# A Lumped Model for Rotational Modes in Phononic Crystals

Pai Peng[1], Jun Mei[2] and Ying Wu[1]

[1]*Division of Mathematical and Computer Sciences and Engineering, King Abdullah University of Science and Technology (KAUST), Thuwal 23955-6900, Saudi Arabia*

[2]*Department of Physics, South China University of Technology, Guangzhou 510641, China*

**Abstract**

We present a lumped model for the rotational modes induced by the rotational motion of individual scatterers in two-dimensional phononic crystals comprised of square arrays of solid cylindrical scatterers in solid hosts. The model provides a physical interpretation of the origin of the rotational modes, reveals the important role played by the rotational motion in the band structure, and reproduces the dispersion relations. The model increases the possibilities of wave manipulation in phononic crystals. In particular, expressions, derived from the model, for eigen-frequencies at high symmetry points unambiguously predict the presence of a new type of Dirac-like cone at the Brillouin center, which is found to be the result of accidental degeneracy of the rotational and dipolar modes.

**Introduction**

Phononic crystals, i.e., structured materials with periodic modulations in their density and elastic coefficients [1-18], have been rapidly developed in recent years due to their applications in manipulating acoustic and elastic waves. Based on their constituents, phononic crystals can be categorized into at least two types. Prominent ones are acoustic crystals, which are solid/fluid inclusions in fluid hosts [1-8]. Another common type is arrays of solid inclusions embedded in solid hosts [9-18]. To differentiate this type from acoustic crystals, we call them elastic phononic crystals (EPC). EPCs are more complex than its acoustic counterparts because of the shear deformation and induced shear-restoring force in solids, features that do not appear in ideal fluids, where here are inviscous liquids or gases [19]. The existence of shear deformation in EPCs is the underlying reason for many salient phenomena of this type of phononic crystal. One such phenomenon is the local rotational motion [12,14,17] in an EPC, which does not exist in an acoustic crystal.

Although rotational motion has been extensively observed in EPCs [12, 14, 17], to the best of our knowledge, it has not yet been modeled. Without a model, it is difficult to accomplish the task of establishing the relationship between rotational modes and the microstructure of an EPC, which is crucial in utilizing the rotational modes to achieve intriguing wave propagation properties with potential applications. In the past, it seems that the focus of work on EPCs was on controlling the usual longitudinal and transverse modes [10,11,13-15] in the crystal rather than on the physics of the rotational modes. Without a clear understanding of the rotational motion, however, it is difficult to fully understand the wave propagation properties in an EPC.

In this paper, we establish a model that describes the rotational modes in two-dimensional EPCs. We show that a simple lumped model captures the essence of the physics of the rotational modes, as well as offers a reasonably good prediction of the band structures of EPCs. Our model reveals the inherent link between the properties of the rotational modes and the microstructure of the EPC, and allows us to engineer an EPC that exhibits an unusual Dirac-like cone [6,20,21] at the Brillouin center. The newly observed Dirac-like cone is obtained by exploiting the rotational motion to achieve accidental degeneracy of a rotational mode and dipolar modes.

**Lumped Model**

The EPC considered in this work is a square array of cylindrical solid inclusions with radii $R$ embedded in another solid matrix. The lattice constant is $a$, which is also the length unit. The elastic wave propagation in such a system is characterized by the displacement of each point in the unit cell. We focus on the wave propagation along the $\Gamma X$ direction and propose the lumped model shown in Fig. 1. $m_1$ and $m_2$ are the masses of the matrix and the scatterer, respectively, and the interaction forces between these masses are modeled by massless "springs" connecting them [22,23]. In fact, these springs are "continuous" springs, but for simplicity and without loss of generality, we demonstrate in Fig. 1 by using two individual springs connected to the scatterer at two arbitrary points, "A" and "B", on the boundary of the scatterer. To study the low frequency rotational modes, where the rotational vibration is more likely to occur in the scatterer about its axis, the mass $m_2$ is considered to be a finite-sized cylinder with radius $R$ and a moment of inertia, $I_2$, rather than a point mass like $m_1$. The finite-size assumption was also used in the study of granular systems [24, 25], where the moment of inertia comes into play. A rotational mode naturally satisfies $\nabla \times \mathbf{u} \neq 0$, which means that the transverse wave is required to excite the rotational motion as $\nabla \times \mathbf{u}_l = 0$ for longitudinal waves. In the low frequency limit, it can be shown that along the $\Gamma X$ direction, only transverse waves are coupled with rotational motion. Therefore, we consider only transverse motion here as the translational movement. As a result, the displacements of the centers of masses are confined along the vertical direction and denoted by $u_{2i-1}$ and $u_{2i}$ for $m_1$ and $m_2$, respectively, where $i$ is the index of the unit cell.

The inter-mass force arises because of the relative distortion between the adjacent masses. There are two sources for the restoring force exerted on $m_1$ in the i-th unit cell. One is the translational movement, i.e., the relative displacement between the centers of the mass and the scatterers nearby, which can be expressed as: $\mathbf{F}_{t\_2i-1} = -K\left[\left(u_{2i-1} - u_{2i}\right) + \left(u_{2i-1} - u_{2(i-1)}\right)\right]\hat{y}$, where $K$ is the spring constant and is related to both the bulk and shear moduli of the matrix [23] . The other source is introduced by the rotational movements of the two neighboring scatterers and takes the form: $\mathbf{F}_{r\_2i-1} = \left[-\frac{K}{\pi}\int_{-\pi/2}^{\pi/2}\theta_{2i}R\cos\varphi_{2i,1}d\varphi_{2i,1} + \frac{K}{\pi}\int_{-\pi/2}^{\pi/2}\theta_{2(i-1)}R\cos\varphi_{2(i-1),2}d\varphi_{2(i-1),2}\right]\hat{y}$, where $\theta$ denotes the rotated angle of the scatterer from its equilibrium position and $\varphi_{2i,1}$ ($\varphi_{2i,2}$) is the polar angle before deformation. Similarly, the total restoring force exerted on $m_2$ can be written as $\mathbf{F}_{t\_2i} = -K\left[\left(u_{2i} - u_{2i-1}\right) + \left(u_{2i} - u_{2i+1}\right)\right]\hat{y}$ and $\mathbf{F}_{r\_2i} = 0$ , which comes from the relative distortion resulting from the translational and rotational movements of the scatterer. The rotation of the scatterer would not generate a net total force on itself, i.e., $\mathbf{F}_{r\_2i} = 0$, but it causes a torque on $m_2$ due to the finite size of $m_2$. Analogous to the resorting force, there are also two sources for the torque, which can be expressed as

$$\mathbf{M}_{t\_2i} = \left[ \frac{KR}{\pi} \int_{-\pi/2}^{\pi/2} (u_{2i} - u_{2i-1}) \cos\varphi_{2i,1} d\varphi_{2i,1} - \frac{KR}{\pi} \int_{-\pi/2}^{\pi/2} (u_{2i} - u_{2i+1}) \cos\varphi_{2i,2} d\varphi_{2i,2} \right] \hat{z} \quad \text{for the}$$

translational part, and $\mathbf{M}_{r\_2i} = -2K(\theta_{2i}R)R\hat{z}$ for the rotational part.

Newton's second law gives us the equations of motion for $m_1$ and $m_2$:

$$\begin{aligned} m_1\ddot{\mathbf{u}}_{2i-1} &= \mathbf{F}_{t\_2i-1} + \mathbf{F}_{r\_2i-1} \\ m_2\ddot{\mathbf{u}}_{2i} &= \mathbf{F}_{t\_2i} + \mathbf{F}_{r\_2i} \\ I_2\ddot{\boldsymbol{\theta}}_{2i} &= \mathbf{M}_{t\_2i} + \mathbf{M}_{r\_2i} \end{aligned} \quad (0)$$

If we suppose that the vibration is time harmonic and invoke the Bloch theorem, we can transform Eq. (0) into the following secular equation:

$$\det \begin{vmatrix} m_1\omega^2 - 2K & 2K\cos(k_x a/2) & -\frac{4}{\pi}iKR\sin(k_x a/2) \\ 2K\cos(k_x a/2) & m_2\omega^2 - 2K & 0 \\ \frac{4}{\pi}iKR\sin(k_x a/2) & 0 & I_2\omega^2 - 2KR^2 \end{vmatrix} = 0, \quad (1)$$

where $k_x$ is the Bloch wave vector along the $\Gamma X$ direction. By solving Eq. (1), we establish the dispersion relations, i.e., $\omega(k)$, for which the expression is complicated in general except for those at the high symmetry points. For example, at the $\Gamma$ point ($k_x = 0$), Eq. (1) gives three eigenfrequencies in terms of the material parameters: $\omega_{1\Gamma} = 0$, $\omega_{2\Gamma} = \sqrt{2KR^2/I_2}$, $\omega_{3\Gamma} = \sqrt{2K(m_1+m_2)/(m_1 m_2)}$. The corresponding results at the X point of the reduced Brillouin zone, ($k_x = \pi/a$) are: $\omega_{1X} = \sqrt{2K/m_2}$, $\omega_{2X} = \sqrt{2K(R^2/I_2 - \Delta)}$ and $\omega_{3X} = \sqrt{2K(1/m_1 + \Delta)}$, where $\Delta$ is a positive quantity and equals $\left( \sqrt{(I_2 - R^2 m_1)^2 + 16 m_1 I_2 R^2 / \pi^2} - |I_2 - R^2 m_1| \right) / 2m_1 I_2$.

**Results and Discussion**

**1) Band structure and eigenstates**

We test the validity of the lumped model by using a real EPC: a square array of steel cylinders with radius $R = 0.2a$ embedded in epoxy. The mass density and longitudinal and transverse velocities inside the epoxy are, respectively, $\rho_1 = 1180$ kg/m$^3$ $c_{l1} = 2540$ m/s and $c_{t1} = 1160$ m/s. The corresponding parameters of the steel are $\rho_2 = 7900$ kg/m$^3$, $c_{l2} = 5800$ m/s and $c_{t2} = 3200$ m/s. We use COMSOL Multiphysics, a commercial package based on the finite-element method, to compute the band structure of this EPC and plot the results in circles in Fig. 2(a). The solid circles highlight the transverse and rotational branches of interest, and the open circles represent the longitudinal modes. The band structure can also be evaluated from the lumped model, i.e. the solution of Eq. (1), in which the masses are $m_1 = \rho_1 a(a - 2R)$ and $m_2 = \rho_2 \pi R^2$. Here, $m_1$ is chosen to be the mass of the portion of the matrix that is sandwiched between two adjacent scatterers, which, as is shown later, effectively contributes to the transverse motion. In this case, $m_2$ is about 1.4 times that of $m_1$. Consequently, the previously derived $\omega_{2\Gamma}$ is smaller than $\omega_{3\Gamma}$. To solve Eq. (2), we need to know the spring constant, $K$, which should be determined from the moduli of the materials, but the relation is complex and beyond the scope of this work. Nevertheless, we can fit its value by choosing *only* one point from the band structure. Here, for simplicity, we adopt the eigenmode at the $\Gamma$ point on the second band and let its frequency equal the lumped model prediction, $\omega_{2\Gamma}$. Thus, the value of $K$ can be evaluated from $\omega_{2\Gamma} = \sqrt{4K/m_2}$. With $m_1$, $m_2$ and $K$, it is straightforward to compute the dispersion relations from the lumped model. The results are plotted in Fig. 2(a) in solid red curves, which agree well with the solid circles, verifying the validity of the lumped model for the lowest three transverse and rotational branches.

To gain a deeper understanding of the highlighted branches, we plot in Figs. 2(b)-2(g) the displacement field patterns associated with six eigenstates at high symmetry points of the Brillouin zone. Color represents the normalized magnitude, with dark red and dark blue corresponding to the maximium (normalized to 1) and zero, respectively, and the small arrows indicate the direction of the displacement. The motions of the cylinder and the matrix can be qualitatively described by the thick arrows. The upper panel corresponds to states at the $\Gamma$ point and the lower panel shows those at the $X$ point. From left to right, the frequency is increasing. Figures 2(b) and 2(c) show translational movement along the direction perpendicular to the wave vector **k**, and imply that the lowest band is a typical acoustic-type branch [26], in which the scatterer and the matrix move as a whole in the zero-frequency limit and the movement is concentrated on the scatterer at the Brillouin zone boundary. Different from Figs. 2(b) and 2(c), pure rotation of the scatterer is found in Fig. 2(d), which is in accordance with the lumped model as frequency $\omega_{2\Gamma}$ is the natural angular frequency of an oscillating rotating cylinder and is a function only of the moment of inertia. The mode on the same band but at the Brillouin zone boundary is exhibited in Fig. 2(e), where the rotational motion of the scatterer is preserved but is weak compared with the strong translational displacements of the matrix. The matrices on the left and right sides are moving in the opposite directions as if they drive the rotational motion of the scatterer. We call this type of mode "in-phase", which is a contrast to the mode plotted in Fig. 2(g), where the rotation of the scatterer is not in line with the translational displacement of the matrix and we call it the "out-of-phase" mode. Figure 2(f) demonstrates a mode that has a similar pattern of the "optical-type" mode [26], where the scatterer and the matrix are moving both vertically but in opposite directions. This, again, agrees with the lumped model prediction, which gives the typical frequency, i.e., $\omega_{3\Gamma}$, of an optical mode of a diatomic chain [26]. In fact, this mode is the transverse state of the two degenerated dipolar modes. The other one depicted in Fig. 2 (a) by the open circles near frequency $\omega_{3\Gamma}$ is longitudinal with the matrix and the scatterer moving horizontally, which is not considered here. Figures 2(d)-2(g) indicate that the matrix above or below the scatterer contributes very little to the transverse movement so that the choice of $m_1$ is justified. However, this is not true for Figs. 2(b) and 2(c), which might be the reason for the discrepancy between the lumped model and the numerical simulation for the lowest branch.

Figure 2(a) shows a band gap between the second and the third band. The lumped model suggests that the width of the gap is determined by the difference between the two masses, i.e., $\Delta\omega = \omega_{3\Gamma} - \omega_{2\Gamma} = \sqrt{2K}\left(\sqrt{1/m_1} - \sqrt{1/m_2}\right)$. For the EPC we have just studied, $m_1$ is smaller than $m_2$. The rotational mode is therefore located below the optical-type mode at the Γ point. If $m_1$ is greater than $m_2$, these modes will interchange their positions. Given the materials, the values of $m_1$ and $m_2$ are simply functions of the sizes of the scatterer. Figure 3(a) shows the band structure of a similar EPC but the radius of the steel cylinder is changed to 0.15a. Good agreement between the lumped model and the numerical calculation is seen again. In this case, $m_1/m_2 \approx 1.45$ and $\omega_{2\Gamma} > \omega_{3\Gamma}$. Figures 3(b) and 3(c) show the field patterns at frequencies of $\omega_{2\Gamma}$ and $\omega_{3\Gamma}$, respectively. Obviously, the one at the lower frequency is an "optical-type" transverse mode degenerating with a longitudinal mode and the other is a rotational mode. This is opposite to the case shown in Fig. 2(a).

**2) Zero gap: a Dirac-like cone**

Since the magnitudes of $\omega_{2\Gamma}$ and $\omega_{3\Gamma}$ depend on the scatterer's size, it is possible to make them equal by carefully tuning the radius of the cylinder. Thus, an interesting result, the zero gap width, is achievable as shown in Fig. 4(a), where the radii of the steel rods are $0.177a$. As expected, we find the rotational mode and the dipolar modes occur simultaneously at a dimensionless frequency $\omega_0 a/2\pi c_{t1} = 0.826$ at the $\Gamma$ point. A bit surprising result is that in the vicinity of this frequency, the dispersion relations become linear as plotted in an enlarged view in Fig. 4(b). An equi-frequency surface at a frequency slightly below $\omega_0$, i.e., $\omega a/2\pi c_{t1} = 0.82$, is plotted in Fig. 4(c), with the blue circles indicating the results calculated by COMSOL. These circles form a perfect and large circle drawn as a red solid curve, which implies that the dispersion relation is isotropic. Similar results are also obtained for the upper branch but are not shown here. The isotropic behavior of the equi-frequency surface somehow contradicts our common understanding that the dispersion relation of an EPC in a square lattice is in general anisotropic [15, 16, 27]. In fact, this special characteristic is a result of accidental degeneracy and ensures that the dispersion relation near the degenerate frequency can be described in terms of a linear cone that intersects a flat sheet in the vicinity of the zone center, which is also called a "Dirac-like" cone and the vertex of the cone is called a Dirac-like point [21]. This behavior looks very similar to the Dirac-like cone of electromagnetic [20] and acoustic waves [6], where the monopolar and dipolar bands meet at the $\Gamma$ point. Dirac-like cones were also found in an EPC when the accidental degeneracy of the dipolar and quadrupolar modes occurs and was understood from an effective medium perspective [14]. In that case, the effective mass density $\rho_{eff}$, and the effective stiffness $1/C_{44}$, are simultaneously zero at the Dirac-like point. Here, for the first time, we discover Dirac-like cones for elastic waves, which is a result of the accidental degeneracy of rotational and dipolar modes. The occurrence of the Dirac-like point cannot be interpreted by conventional effective medium theory [11, 14], however, because the wavelength in the steel rods is longer than that in the epoxy matrix, deeming the conventional effective medium description inapplicable. It is, on the other hand, well predicted by the lumped model and the dispersion relation can also be evaluated by the lumped model as shown by the solid red lines in Figs. 4(a) and 4(b).

Figures 4(d)-4(i) show the displacement field patterns near the Dirac point with a small **k** along $\Gamma X$ direction (left panel) and $\Gamma M$ direction (right panel). Figures 4(d) and 4(e) show the real and imaginary part of the displacement fields of the mode on the lower band (marked as "A" in Fig. 4(b), where $k_x = 0.05\pi/a$, and $k_y = 0$). It is clear that this eigenmode is a linear combination of dipole and rotation excitation. The real part of the displacement is concentrated within the matrix along the direction perpendicular to **k**. Thus, it can be regarded as a transverse component, which is similar to the transverse dipolar mode mentioned previously [as shown in Fig. 2(f)]. Similarly, the imaginary part can be viewed as a rotational component, in which the motion is mainly the rotation of the steel rod. Because the eigenstates of the mode with the same **k** but on the upper branch exhibit the same behavior, we do not plot them here. Figure 4(f) shows the real part of the displacement of the mode on the middle band (marked as "B" in Fig. 4(b)). Interestingly, a longitudinal dipolar

mode with its vibration direction parallel to **k** is found. The imaginary part of mode "B" is two orders of magnitude smaller than its real part and is not presented. The eigenmode along the $\Gamma M$ direction is plotted in Figs. 4(g) and 4(h) for the real and imaginary parts of the mode that belongs to the lower branch of the linear cone (marked as "C" in Fig. 4(b), whose $k_x = 0.05\pi/a$ and $k_y = 0.05\pi/a$). The behavior of the displacement fields is in general similar to the one along the $\Gamma X$ direction, but the rotation and transverse component interchange their positions, i.e., the real part represents a rotation state whereas the imaginary part represents a transverse state. Similar eigenstates are found on the upper band with the same **k**. In the middle branch, the real part of the mode marked as "D" in Fig. 4(b) is plotted in Fig. 4(i), whose vibration direction is parallel to **k**. The imaginary part of this mode is, again, very small compared with the real part. Together, these observations suggest the remarkable properties of this Dirac-like cone: the linear cone is mainly contributed by the hybrid modes of the rotational and transverse components, and the middle branch is longitudinal. This means that the isotropic linear cone can only be coupled with the transverse wave in all directions.

**Conclusions and Outlook**

In this work, we propose a simple lumped model that can account for the local rotation of the scatterer in certain two-dimensional EPCs. The model is useful for a physical understanding of the lowest transverse and rotational bands. It also provides a good prediction of the dispersion relations. With the model, it is convenient to go beyond the translational degree of freedom and utilize the rotational mode to achieve interesting wave propagation properties.

As an example, we demonstrate the occurrence of a new type of Dirac-like cone, which is a result of the accidental degeneracy of the rotational and dipolar mode at the Brillouin zone center. This Dirac-like cone possesses isotropic dispersion relations that can be excited only by transverse waves in the vicinity of the Dirac-like point even for a square lattice, which might lead to interesting elastic wave transport properties.

**Acknowledgments**

The authors would like to thank Professor P. Sheng, Professor Z. Q. Zhang, Professor Z. Y. Liu, and Dr. C. Y. Qiu for discussions. The work described here was supported by the KAUST start-up package and by China's Fundamental Research Funds for the Central Universities (Grant No. 2012ZZ0077).

**Figures**

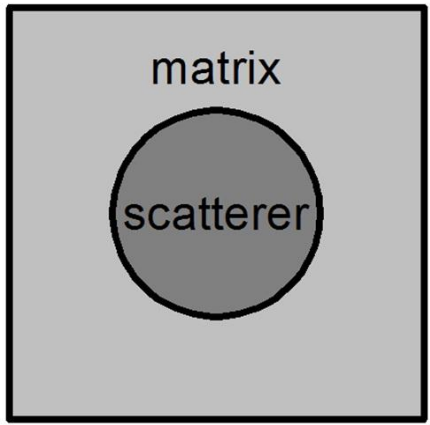 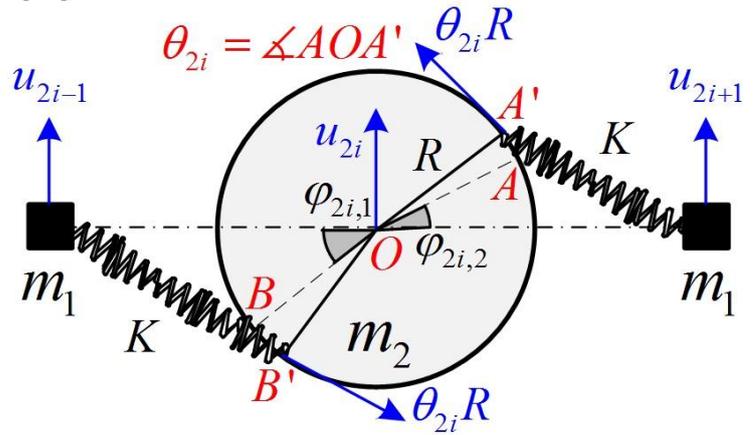

FIG. 1 (a) The unit cell of the phononic crystal. (b) A schematic of the lumped model. The solid black squares indicate the point-mass $m_1$ representing the matrix, and the circles indicate the finite-sized cylindrical scatterer with mass $m_2$. $u$ is the transverse displacement. A and B are arbitrarily chosen points on the boundary of the scatterer, whose equilibrium polar angles are $\varphi_{2i,1}$ ($\varphi_{2i,2}$). They rotate at an angle of $\theta_{2i}$ about the axis of the cylinder from their equilibrium positions to points $A'$ and $B'$. The springs represent the restoring force between two adjacent masses.

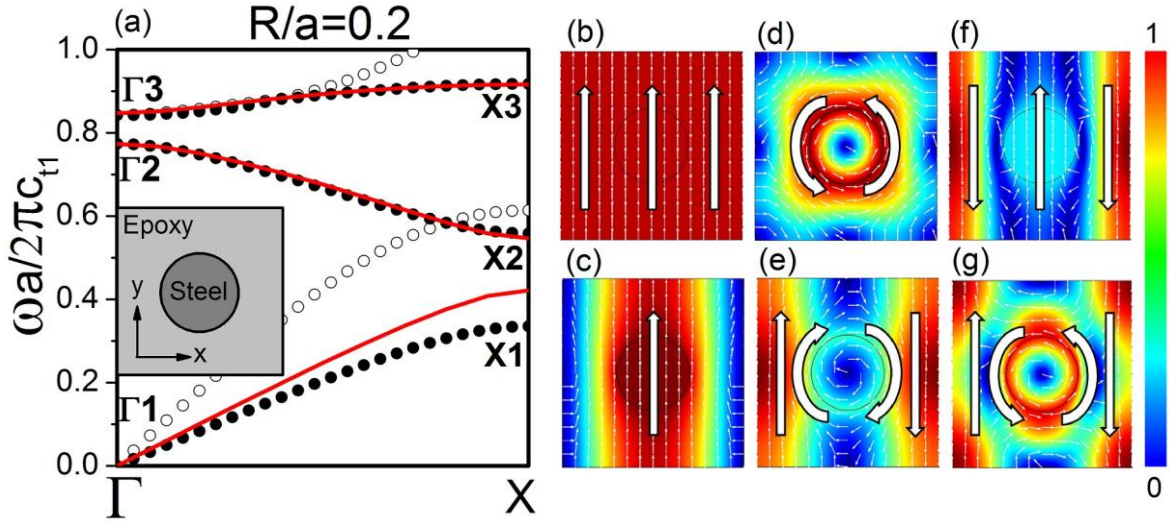

FIG. 2 (a) Band structures of a square array of steel cylinders embedded in epoxy with radius $R/a = 0.2$ The solid dots correspond to the transverse- and rotation-related modes. The red curves indicate the dispersion curves predicted by the lumped model.

(b)-(g) The displacement field distributions of the eigenmodes marked in (a). The correspondences are: (b) $\Gamma 1$, (c) $X1$, (d) $\Gamma 2$, (e) $X2$, (f) $\Gamma 3$, (g) $X3$. Dark red and dark blue corresponds to one and zero of the normalized magnitude, respectively, and thin arrows indicate directions. Thick arrows demonstrate, qualitatively, the movement of the matrix and the scatterer.

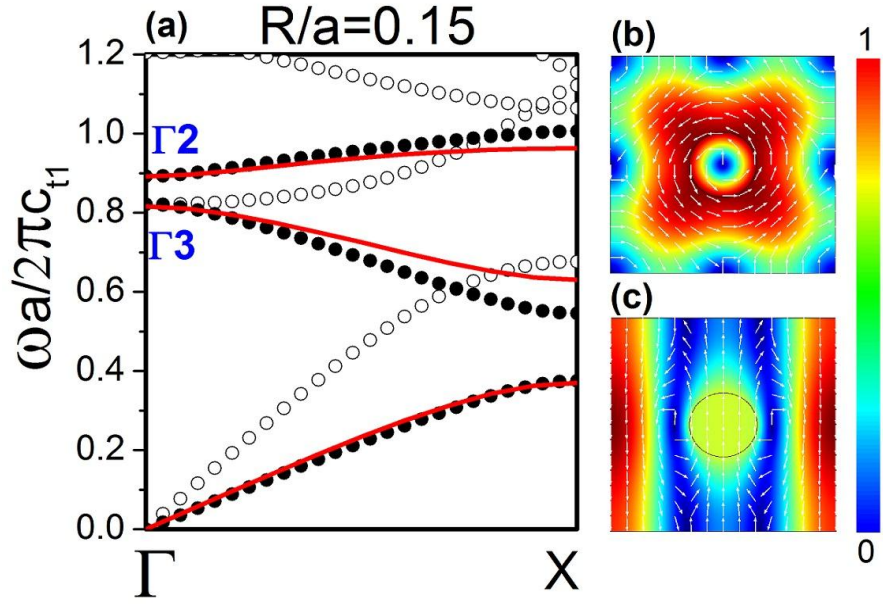

FIG. 3 (a) Band structures of the same system as shown FIG. 2, but the radius is changed to $R/a = 0.15$. (b) and (c) are the displacement field distributions of the eigenstates marked as Γ2 and Γ3, respectively.

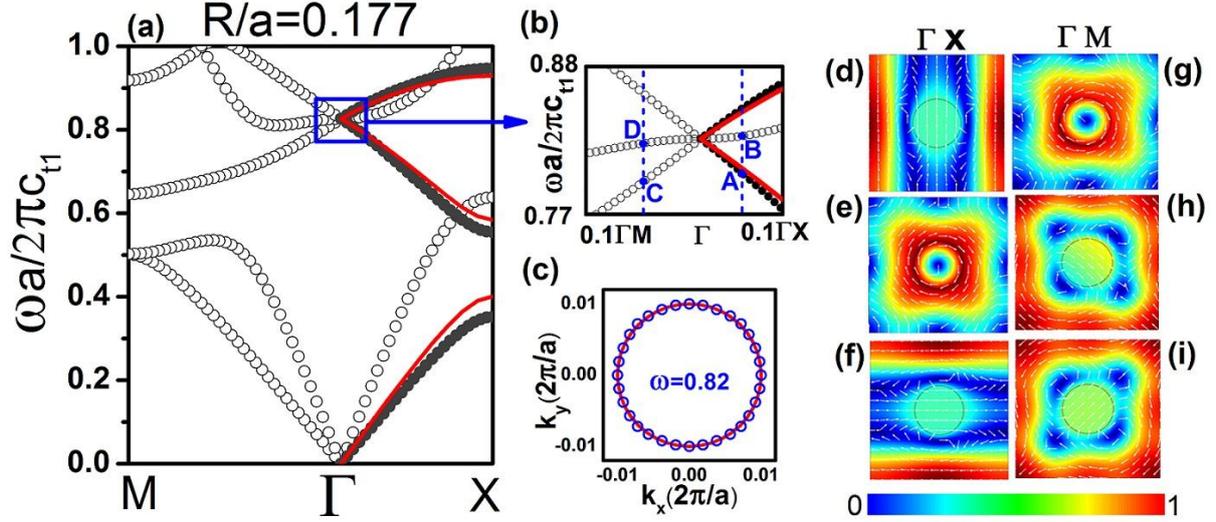

FIG. 4 (a) Full band structures of the same system as shown in FIG. 2, but the radius is changed to $R/a = 0.177$. (b) An enlarged view of the band dispersion near the Dirac-like cone. (c) The equi-frequency surface at the dimensionless frequency $\omega a/2\pi c_{t1} = 0.82$. The blue circles indicate the numerical calculations, which form a perfect circle plotted in red curve. (d)-(f) The displacement field distribution of the eigenmodes near the Dirac-point along the $\Gamma X$ direction: (d) and (e) the real part and the imaginary parts of displacement fields of state "A", (f) the real part of the displacement field of state "B". (g)-(i) corresponding results for modes along the $\Gamma M$ direction. (g) and (h) the real part and the imaginary parts of the displacement field of state "C", (f) the real part of the displacement field of state "D".